\documentclass[12pt]{iopart}
\usepackage{iopams}
\usepackage{graphics}

\begin{document}

\title{Vortices in nonlocal Gross-Pitaevskii equation}

\author{Valery S Shchesnovich$^{1,2}$  and
Roberto A Kraenkel$^{1}$}

\address{$^1$Instituto de F\'{\i}sica Te\'{o}rica,
Universidade Estadual Paulista,
Rua Pamplona 145, 01405-900 S\~{a}o Paulo, Brazil\\
$^2$Departamento de F\'{\i}sica - Universidade Federal de Alagoas, Macei\'o
AL 57072-970, Brazil }

\eads{\mailto{valery@ift.unesp.br}, \mailto{kraenkel@ift.unesp.br}}

\begin{abstract}
We consider vortices in the nonlocal two-dimensional Gross-Pitaevskii
equation with the interaction potential having the Lorentz-shaped
dependence on the relative momentum. It is shown that in the Fourier series
expansion with respect to the polar angle the unstable modes of the axial
$n$-fold vortex have orbital numbers $l$ satisfying $0<|l|<2|n|$, similar
as in the local model. Numerical simulations  show that nonlocality
slightly decreases the threshold rotation frequency above which the
nonvortex state ceases to be the global energy minimum and decreases the
frequency of the anomalous mode of the 1-vortex. In the case of higher
axial vortices, nonlocality leads to the instability against splitting with
creation of antivortices  and gives rise to additional anomalous modes with
higher orbital numbers. Despite new instability channels with creation of
antivortices, for a stationary solution comprised of vortices and
antivortices there always exist another vortex solution, comprised solely
of vortices, with the same total vorticity but with a lower energy.

\end{abstract}


%\keywords{Quantized vortices in condensates, nonlocal interactions, vortex
%stability}

\maketitle
\section{Introduction}

The Gross-Pitaevskii (GP) equation  for the order parameter is derived by
replacing the interaction potential by the Fermi zero-range
pseudo-potential \cite{gross,pitaevskii} (see also the review
Ref.~\cite{revBEC1}). For the gaseous Bose-Einstein condensates (BECs)
usage of the pseudo-potential is justified, since the ``gaseousness
parameter'', $\varrho a_s^3$, where $\varrho$ is the particle density and
$a_s$ is the scattering length, is of order $10^{-3}$. In a cold dilute gas
only binary collisions at low energy are relevant and these collisions are
characterized by a single parameter, the s-wave scattering length,
independently of the interaction  potential.

The predictions of the  mean-field theory agree with the experimental results
on BEC in dilute gases when both the quantum fluctuations and the cloud of
non-condensed atoms can be neglected. In connection with gaseous BECs, the
predictions of the GP equation were first analyzed in Ref.~\cite{BPtheor},
where insightful estimates were given and important scaling relations were
established. The mean-field theory was subsequently employed for account of
the  dynamical properties of BEC in ($i$) description \cite{expannum} of the
ballistic expansion of BEC \cite{exp1} and ($ii$) quantitative theoretical
account \cite{excitnum} of the condensate excitation spectra  observed in
Ref.~\cite{excitexp}. In both cases the agreement with the experiment was
within 5\% without any fitting parameters. The existence of the macroscopic
phase in BEC, which is one of the premises for introduction of the order
parameter, was  experimentally verified in the interference picture of two
BECs in a double-well potential \cite{interexp}. Importantly, the
quantitative results of this experiment were reproduced within the GP-based
mean-field theory \cite{interth}.

The correspondence between  theory and experiment was enhanced dramatically
with the observation of quantized vortices. Vortices are experimentally
created in BEC using various techniques of orbital momentum transfer to the
condensate, such as ``phase imprinting'' in a two-component condensate
\cite{expvrt1}, using a laser beam stirrer~\cite{expvrt2,expvrt3,expvrt3a},
through the decaying solitons into vortex rings \cite{solitvrt1,solitvrt2},
rotation of the magnetic trap~\cite{expvrt4}, and rotation of the thermal
cloud during the evaporating cooling process~\cite{expvrt5}. Recently,
coreless vortices were produced by the phase imprinting method in a spinor
condensate \cite{coreless1}.

The mean-field theory with account for the mechanism of the angular momentum
transfer to the condensate describes the experiments on vortices in gaseous
BECs. First of all the vortex state must have lower energy in the rotating
frame than the condensate without vortices. This picture resulted in the
first proposed critical rotation frequency $\Omega_v$ \cite{OmegEner} for
observation of vortices. However, it was shown later that the nonvortex state
remains a local minimum at the rotation frequencies higher than $\Omega_v$
\cite{phasdiagr} and, hence, an energy barrier may prevent nucleation of
vortices. Also, the vortex precession around the center of the trap was
identified with its anomalous mode in Ref.~\cite{anomalmode}. A surface mode
version of the Landau theory was then suggested as a mechanism for vortex
nucleation \cite{surfmode1} resulting in a much higher critical nucleation
frequency, defined as a minimum over all resonances with surface modes of
orbital numbers $l$ and frequencies $\omega_l$: $\Omega_{c}= \mathrm{min}
(\omega_l/l)$. The surface mode theory is generally consistent with the
experiments \cite{expvrt1,expvrt2,expvrt3,expvrt4}. The dynamic vortex
formation theory in which rotating atomic cloud plays an essential role was
proposed in Refs.~\cite{vrtnuclcloud1,thermcloud}. The general
vortex-nonvortex state stability diagram for finite temperatures was obtained
in Ref.~\cite{stabdiagrfint}. The surface mode spectrum at finite
temperatures was computed in Ref.~\cite{surfatfint} within the Popov version
of the Hartree-Fock-Bogoliubov theory. It was found that the thermodynamic
frequency $\Omega_v$ and the critical nucleation frequency $\Omega_c$ of the
surface-mode theory increase with the temperature (however, the nucleation
frequency due to the quadrupole mode remains almost unaffected). In
Refs.~\cite{numer1,numer2} extensive numerical simulations of the GP equation
and the Bogoliubov linear equations in spherically and cylindrically
symmetric traps were performed, where the critical vortex nucleation
frequencies were found.

The excitation of the  surface modes is not the only possible mechanism for
vortex nucleation. Action of a localized perturbation, a small stirrer, can
allow for vortex formation  below the nucleation threshold $\Omega_c$ of
the surface-mode theory. In the  experiment  of Ref.~\cite{expvrt3a} a
nearly pure condensate was excited  by a stirrer with the variable
characteristic size. The rapid nonresonant vortex formation was observed
for small stirrer size at the rotation frequency lower than $\Omega_c$,
indicating on a local mechanism related to the formation of
vortex-antivortex pairs (see also Ref.~\cite{localnucl1}). In
Ref.~\cite{localnucl2} a classical solid-body model for the angular
momentum buildup was shown to describe the above experiment when the
density of vortices is high. On the other hand, it is known that in
effectively two-dimensional BEC the vortex-antivortex pair formation is the
dominant mechanism of the energy transfer from a moving object to the fluid
at velocities above the threshold value \cite{vortform,vortshed}. In 3D
geometry  the vortex rings are formed instead and, depending on the drag
force, the object is either slowed down or gets trapped inside the ring
core \cite{vortring}.

In connection of applicability of the GP equation, an important mathematical
result was also established: the Gross-Pitaevskii functional is the
asymptotic limit for gaseous BECs of the N-particle energy functional both in
three \cite{deriv3D} and in two \cite{deriv2D} dimensions (see also
Ref.~\cite{derivnew}).

The concept of  quantized vortices first appeared in connection with liquid
helium in the pioneering  ideas of Onsager and Feynman \cite{OnsgFeynm}
(see also Refs.~\cite{vrtHe,vrtdyn,hist}). Quantized vortices were
experimentally detected in liquid helium \cite{Vinen}. Similar to gases,
single vortices and vortex arrays are observed in helium II (for instance,
via the ion trapping technique \cite{photvrtHe1,photvrtHe2}), this subject
is reviewed in Ref.~\cite{vrtHe}.

Liquid helium,   in contrast to cold alkali gases, is a dense strongly
correlated system (the parameter $\varrho a_s^3$ is of order 0.1).   In
helium the interatomic distance is on the order of the interaction range
(see, for instance, table I in Ref.~\cite{FHerev}).  The strong dissimilarity
between the two systems manifests itself in the vortex core size: the vortex
core in gases is much larger than the average interatomic distance, while in
liquid helium it is of the same order. Evidently, there is no justification
for neglecting the range of interaction in description of the macroscopic
properties of liquid helium.

Several mean field functionals for  the superfluid helium are
discussed in literature \cite{FHe1,FHe2,newhe1,FHe3,FHe4} (consult
also the review Ref.~\cite{FHerev}). The simplest form of the
correlation energy part for liquid helium is as
follows~\cite{FHe1}
\begin{equation}
\mathcal{E}_c[\varrho] =
\int\mathrm{d}^3\mathbf{r}\left\{\frac{b}{2}\varrho^2 +
\frac{c}{2}\varrho^{2+\gamma} + d(\nabla \varrho)^2\right\},
\label{corrlenerg}\end{equation}
 where $\varrho $ is the local density, and $b$, $c$, and $\gamma$
are the phenomenological parameters. Note that the last term in
(\ref{corrlenerg}) accounts for the nonlocal  interactions in liquid helium
and is crucial for the correct  description of its macroscopic properties
\cite{FHe2,FHe3}. Instead of the gradient nonlocality as in
(\ref{corrlenerg}), an {\it ad hoc} potential with fitting parameters was
also proposed~\cite{FHe4}. The higher-order term $\varrho^{2+\gamma}$ is
essential for counteraction of the non-physical mass concentrations in a
nonlocal model with attractive-repulsive interactions \cite{highorder}.
Thus, complicated density functionals are proposed to explain the
properties of liquid helium. However, it has been known for quite some time
that some aspects of vortex dynamics in liquid helium, such as the
annihilation of vortex rings \cite{Heloc1}, the nucleation of vortices
\cite{Heloc2}, the vortex line reconnection \cite{Heloc3}, and the
superfluid turbulence \cite{Heloc4} are in fact captured by the local GP
model.

In this paper we study the effect  of the finite-range of interaction on
vortices. Our motivation is that, on one hand, some properties of vortices
in the strongly interacting systems seem to be described by the local model
and, on the other hand, since vortices appear due to balance between the
dispersion and nonlinearity, modification  of the interaction potential may
lead to different stability properties of vortices. Thus our goal is to
find out what changes in the stability properties of vortices can be
attributed solely to nonlocality.

Vortices in a nonlocal model were studied recently in Refs.~\cite{FHe4,FHe5},
where a phenomenological potential was used which brought into agreement the
vortex core size with the healing length and allowed to reproduce the Landau
phonon-roton dispersion curve. Our nonlocal model is obtained by substitution
of the Fermi zero-range potential by the Yukawa finite-range repulsive
potential (the Macdonald function in two dimensions). As we study only the
effect of the interaction range, we adopt the simplest model of a nonlocal
interaction potential without the attractive part. Thus, we can drop the
higher-order nonlinear term used to compensate collapse due to attraction. To
facilitate the comparison with the local theory we will keep the external
confining potential in our nonlocal GP equation (the effects solely due to
nonlocality are independent of the external potential). For reasons of
accuracy of the numerical analysis, we restrict ourselves to the
two-dimensional model (which corresponds to oblate geometry of the system).

The  nonlocal GP  theory was also recently used to study the effect of
nonlocality of the attractive atomic interactions on stability of the ground
state against collapse.  It is found that nonlocality suppresses
collapse~\cite{negnonloc1} and is responsible for appearance of stable
self-trapped configurations~\cite{negnonloc2,negnonloc3}.

The paper is organized as follows. In section \ref{Model2D} we argue that
the simplest finite-range interaction  potential is the Yukawa potential
(in 3D) and find the corresponding potential in 2D. Then, in section
\ref{Stability}, we show that the unstable orbital modes of the axial
$n$-fold vortex  have orbital numbers $l$ satisfying $0<|l|<2|n|$, while
the ground state is spectrally stable -- precisely as in the local theory.
In section~\ref{Numerics} we first numerically study the effect of the
interaction range on  the critical thermodynamic frequency $\Omega_v$ and
the stabilization frequency of the 1-vortex. Then we consider the effect of
nonlocality on the stability properties of axial 2- and 3-vortices and look
for the vortex-antivortex solutions corresponding to the $n$-vortex
splitting instabilities caused by the nonlocality for various values of the
vorticity $n$. Section \ref{Summary} contains concluding remarks.

\section{Nonlocal Gross-Pitaevskii model}
\label{Model2D}

The Fermi pseudo-potential (given by the Dirac distribution) has one
parameter -- the scattering length. Thus, the simplest model which can
describe nonlocal repulsive interactions involves introduction of an
effective potential with two parameters: the strength of interaction and
the interaction range. The symmetry property and assumption of the
short-range interaction allow us to single out the effective potential with
two parameters. Indeed, the contribution to the energy functional due to
the atomic interactions reads
\begin{equation}
\mathcal{E}_{int} =
\frac{g}{2}\int\mathrm{d}^3\mathbf{r}\int\mathrm{d}^3\mathbf{r}^\prime
|\Psi(\mathbf{r})|^2K(\mathbf{r}^\prime,\mathbf{r})|\Psi(\mathbf{r}^\prime)|^2.
 \label{gennonl}\end{equation}
Here $g$ is the strength of interactions and the kernel is normalized as
$\int\mathrm{d}^3\mathbf{r}K(\mathbf{r}) = 1$. For the spherically symmetric
interaction the kernel $K$ depends only on the relative distance: $K =
K(|\mathbf{r}^\prime - \mathbf{r}|)$. Its Fourier image is then a real
function of the squared wave number $\mathbf{k}$: $\hat{K} =
\hat{K}(\mathbf{k}^2)$. In the zero-range limit (the Fermi pseudo-potential)
$\hat{K}=1$. In the next order of the approximation we have $\hat{K} = 1 -
\epsilon^2\mathbf{k}^2$, where $\epsilon$ is the effective interaction range.
However, in this form, the expansion has a problem when substituted into the
Fourier integral over all $\mathbf{k}$, since the second term is unbounded
(the correction would be the dominant term). A remedy for this is to adopt an
equivalent expression (up to the next order in $\epsilon$) -- the Lorentzian:
\begin{equation}
\hat{K} = \left(1 + \epsilon^2\mathbf{k}^2\right)^{-1}.
\label{lorenz}\end{equation}

The Lorentzian has been used recently as an approximation of the finite-range
potential in Refs.~\cite{negnonloc2,negnonloc3}.

Though we will use the Lorentzian in our numerical simulations, the analytical
approach presented in this paper is valid for a certain class of interaction
potentials to be specified in the next section. The Gaussian
\begin{equation}
\hat{K} = C_\epsilon\exp\left(-\epsilon^2\mathbf{k}^2\right),
\label{gauss}\end{equation}
which is another common choice for the phenomenological interaction potential,
belongs to our class. In the limit of short-range interactions  all the potentials
of our class can be approximated by the Lorentzian (\ref{lorenz}).

In  three dimensions the Lorentzian corresponds to the Yukawa effective potential
($a\equiv \epsilon$)
\begin{equation}
 K_{3D}  = \frac{1}{4\pi a^2 r}\exp\left(-\frac{r}{a}\right),
\label{yukava}\end{equation}
while in two dimensions the effective interaction potential is given by
\begin{equation}
 K_{2D}  = \frac{1}{2\pi a^2}K_0\left(\frac{r}{a}\right),
 \label{K2D}\end{equation}
where $K_0(z)$ is the Macdonald function (see, for instance,
Ref.~\cite{Bessel}). The kernels (\ref{yukava}) and (\ref{K2D}) reduce to
the Dirac distribution in the limit of the zero-range interaction:
\begin{equation}
\lim_{a\to 0}K_{3D}(\mathbf{r}) = \delta_{3D}(\mathbf{r}),\quad \lim_{a\to
0}K_{2D}(\mathbf{r}) = \delta_{2D}(\mathbf{r}).
\end{equation}

Both kernels (\ref{yukava}) and (\ref{K2D}) have {\it integrable}
singularity at $r = 0$. This is evident in the case of the Yukawa kernel,
whereas for the 2D-kernel the following asymptotics holds  $K_0(x) \sim
\ln(2/x)$ as $x\to 0$ \cite{Bessel}. Hence, due to $\ln(x) =
o(x^{-\alpha})$ for all $\alpha>0$ as $x\to0$, the integrability property
follows.

In the following we will also use the operator form, given as $\Lambda^{-1}$,
for the effective potentials (\ref{yukava})-(\ref{K2D}), which then become
the Green functions for the positive-definite operator $\Lambda$:
\begin{equation}
\Lambda = 1 - a^2\nabla^2.
\label{Lambda}\end{equation}

The corresponding nonlocal GP equation  (in the reference frame rotating with
the frequency $\Omega$) reads
\begin{equation}
i\hbar\partial_t \Psi = -\frac{\hbar^2}{2m}\nabla^2\Psi + V\Psi - \Omega
L_z\Psi + g\Psi\Lambda^{-1}|\Psi|^2. \label{GP3D}\end{equation} Here $L_z$
is the angular momentum projection on the $z$-axis: $L_z =
-i\hbar(x\partial_y - y\partial_x)= -i\hbar\partial_\theta$, where $\theta$
is the polar angle in the transverse dimensions. To facilitate the
comparison with the local model we will use the parabolic external
potential confining in the transverse dimensions:
\[
V = \frac{m\omega_\perp^2}{2}{\rho}^2 = \frac{V_0}{a_\perp^2}\rho^2,
\]
where $\brho \equiv (x,y)$ and $\rho = |{\brho}|$, $V_0 =
\hbar\omega_\perp/2\ $ is the characteristic energy of the trap, and the
characteristic length scale is $a_\perp = \sqrt{\hbar/(m\omega_\perp)}$.
For simplicity, we consider the periodic boundary conditions along the
$z$-axis, $\Psi(z+d) = \Psi(z)$.

In the following it will be convenient to use the dimensionless variables:
\begin{equation}
\fl t^\prime = \frac{\omega_\perp}{2}\,t,\quad \mathbf{r}^\prime =
\frac{\mathbf{r}}{a_\perp},\quad \psi = \frac{a_\perp
\sqrt{d}}{\sqrt{N}}\Psi,\quad \Omega^\prime =
\frac{2\Omega}{\omega_\perp},\quad L_z^\prime = \frac{L_z}{\hbar},\quad
g^\prime = \frac{gN}{V_0 a_\perp^2 d},
 \label{rescaledvar}\end{equation} Here
$N$ is the number of atoms in the condensate per $z$-period $d$. Note that
the wave function $\psi$ is normalized to 1.   The interaction range is also
normalized by the characteristic length of the transverse potential $a^\prime
= a/a_\perp$. Below we will use only the dimensionless form of equation
(\ref{GP3D}), thus we drop the primes in all the variables.

Assuming no dependence on $z$, the axially symmetric $n$-fold vortex solution reads
$\psi = e^{in\theta-i\mu t}A(\rho)$ (here $A$ depends on $n$), with the amplitude
$A$ satisfying
\begin{equation}
L_0 A \equiv \left\{ -\nabla^2_\rho + \frac{n^2}{\rho^2} + V(\rho) + gF_0 - (\mu
+\Omega n) \right\}A = 0,
\label{eqAn}\end{equation}
\begin{equation}
F_0\equiv \left(1 - a^2\nabla^2_\rho\right)^{-1}A^2,
 \label{eqF0Lorentz}\end{equation}
here $\nabla^2_\rho \equiv \rho^{-1}\partial_\rho\rho\partial_\rho$ is the
radial part of the Laplacian in 2D and
$2\pi\int_0^\infty{\rho}\mathrm{d}\rho A^2 = 1$.

\section{Stability of the axial vortices in nonlocal interactions}
\label{Stability}

The stability properties  of the  axial $n$-fold vortices in the local GP
equation are well known. In BEC with repulsive interactions the unstable
orbital modes $\sim e^{il \theta}$  of the $n$-vortex in the axisymmetric
trap have orbital numbers $l$ satisfying  $0 < |l|<2|n|$ \cite{stab1} ($l$
is the angular momentum  of the mode). The 1-vortex has at least one
anomalous mode (exactly one in 2D), i.e., the mode with positive norm and
negative energy, qualitatively predicted in Ref.~\cite{stab2} and found
later in Ref.~\cite{anmmode}. The number of anomalous modes of the 1-vortex
depends on geometry of the trap: in a prolate trap additional anomalous
modes are possible \cite{anomalmode,numer2}, which describe bending of the
vortex lines. The detailed phase diagram for vortex stability in 2D was
numerically obtained in Ref.~\cite{phasdiagr,stabdiagrfint}. The role of
the excitations bearing negative energies on  vortex instability was
clarified in Ref.~\cite{rolenegenr}. Finally, it was numerically found that
there are intervals of the interaction strength $g$ where the axial
$n$-vortices with $n\ge2$ are dynamically unstable \cite{AxialStab}.

These results concern condensates with repulsive interactions. It was also
shown that in the {\it attractive} condensate the 1-vortex solution
undergoes splitting due to unstable quadrupole mode \cite{attrinstab}.

Related issues of vortices in the Ginzburg-Landau equation (which represents
nonconservative generalization of the GP equation) were considered in
Refs.~\cite{stabGL2D,stabGL3D1,stabGL3D2} and the properties of the optical
vortices are reviewed in Ref.~\cite{optvrtstab}.

Here we show that there is a class of interaction potentials for which the
unstable modes of the $n$-fold axial vortex have orbital numbers in the
interval: $0< |l|<2|n|$, i.e., exactly  as in the local model.

It is convenient to use another form of the nonlocal GP equation with an
additional dependent variable $F$  describing the nonlinear term (below all
equations are in 2D):
\begin{equation}
\left(1 - a^2\nabla^2\right)F = |\psi|^2,
\label{Fdef}\end{equation}
\begin{equation}
\left\{-i\partial_t -\nabla^2 + V(\rho) - \Omega\tilde{L}_z +
gF\right\}\psi = 0.
 \label{GP2DF}\end{equation}
Here  $V = V(\rho)$ is the general confining trap which allows for the
axial vortex solutions with the amplitude decreasing to zero as
$\rho\to\infty$.

The class of interaction potentials  for which  the main result below is
valid is defined as follows. The general expression for the nonlinear term
$F$, corresponding to the general kernel $K$, reads
\begin{equation}
F = \int\mathrm{d}^2\brho^\prime K({\brho},{\brho}^\prime)
|\psi({\brho}^\prime)|^2.
 \label{genK}\end{equation}
Vortex solutions are possible for the kernels  $K = K(|{\brho}^\prime -
{\brho}|)$. Indeed, the quantity $F_0$ (a generalization of that in
(\ref{eqAn})) is given by the following integral
\begin{equation}
F_0=\int\limits_0^\infty\rho^\prime\mathrm{d}\rho^\prime\left\{\int\limits_0^{2\pi}\mathrm{d}\theta^\prime
K([\rho^2+{\rho^\prime}^2 -
2\rho\rho^\prime\cos(\theta^\prime-\theta)]^{\frac{1}{2}})\right\}A^2(\rho^\prime).
 \label{F0gen}\end{equation}
Obviously, $F_0$ is a function of $\rho$ only and this fact allows, in
principle, existence of the axial vortex solutions.

We consider the class of the interaction potentials of the type $K =
K(|{\brho}^\prime - {\brho}|)$ for which the coefficients of the operator
$\Lambda^{-1}$ in the Fourier expansion with respect to the polar angle, defined as
\begin{equation}
\fl \Lambda^{-1}_lf \equiv
\int\limits_0^\infty\rho^\prime\mathrm{d}\rho^\prime\left[\int\limits_0^{2\pi}\mathrm{d}\phi
\,\cos(l\phi)K([\rho^2+{\rho^\prime}^2-2\rho\rho^\prime\cos\phi]^{\frac{1}{2}})
\right]f(\rho^\prime),\quad l=0,1,2,\ldots,
\label{Lambdgenell}\end{equation}
i.e., the radial operators with the kernels given by the expression in the square
brackets in (\ref{Lambdgenell}), are all positive definite. For example, the
Gaussian interaction potential
\begin{equation} K = \frac{1}{2\pi
a^2}\exp\left(-\frac{\rho^2}{2a^2}\right)
\label{gausskern}\end{equation}
 possesses this property. Indeed, the radial operator
$\Lambda^{-1}_l$ is given by the expression
\begin{equation}
\Lambda^{-1}_lf
=\frac{1}{a^2}\int\limits_0^\infty\rho^\prime\mathrm{d}\rho^\prime
\exp\left(-\frac{{\rho^\prime}^2+\rho^2}{2a^2}\right)I_l
\left(\frac{\rho\rho^\prime}{2a^2}\right)f(\rho^\prime),
\label{gausLambdell}\end{equation} where $I_l(x)$ is the Bessel function of
the second kind. There is a sufficiently large $R$ such that the sign of
the quadratic form of the operator $\Lambda_l^{-1}$ on some $f=f(\rho)$ is
given by the sign of the finite integral
\[
\int\limits_0^R\int\limits_0^R\mathrm{d}\rho\mathrm{d}\rho^\prime
\chi(\rho)I_l\left(\frac{\rho\rho^\prime}{2a^2}\right)\chi(\rho^\prime),
\]
with $\chi\equiv \exp\{-\rho^2/2a^2\}\rho f(\rho)$. Since the Taylor expansion of
the modified Bessel function  $I_l(\rho\rho^\prime/2a^2)$ is a uniformly convergent
series in powers of $\rho\rho^\prime$  with positive coefficients, the positivity
of the operator $\Lambda_l$ (\ref{gausLambdell}) follows.

For the  interaction potential (\ref{K2D}), corresponding to the Lorentzian,  the
operator $\Lambda_l$ is given by
\begin{equation}
\Lambda_l = 1 - a^2\left[\nabla^2_\rho - \frac{l^2}{\rho^2}\right].
 \label{Lambdaell}\end{equation}
Obviously, the radial operators $\Lambda_l$, $l=0,1,2,...$, are positive
definite  with respect to the inner product defined by the integral
$\int_0^\infty\rho\mathrm{d}\rho(\cdot)$.

We assume that the system (\ref{eqAn}) and (\ref{F0gen})  with a given
axisymmetric external potential $V(\rho)$ admits axial $n$-vortex
solutions. \smallskip

\noindent \textbf{Definition}  {\sl We will say that the interaction
potential with the kernel $K(|{\brho}^\prime-{\brho}|)$ satisfies the
positivity property if  the associated radial operators $\Lambda^{-1}_l$
(\ref{Lambdgenell}), $l=0,1,2,3,...$, are positive definite.}
\medskip

\noindent {\textbf{Theorem}  {\sl The axial $n$-vortex solution to the system
(\ref{GP2DF}) and (\ref{genK}) with the interaction potential satisfying the
positivity property can be spectrally unstable only with respect to
perturbations which have non-zero projection on the orbital basis
$e^{il\theta}$  in the interval $0<|l|<2|n|$. The ground state solution is
spectrally stable.}
\medskip

\noindent {\sl Proof.} Let us linearize the system
(\ref{GP2DF})-(\ref{genK}) about the axial $n$-vortex solution. Expanding
the linear corrections $\Phi$ and $F_1$,
\begin{equation}
\psi = \{A(\rho) + \epsilon\Phi(\rho,\theta,t)\}e^{in\theta - i\mu t},\quad F =
F_0(\rho) + \epsilon F_1(\rho,\theta,t),
\label{perturba}\end{equation} in the
Fourier series with respect to $\theta$,
\begin{equation}
\Phi = \sum_{l\ge0} u_l(\rho,t)e^{il\theta} +
v^*_l(\rho,t)e^{-il\theta},\quad F_1 = \sum_{l\ge0}b_l(\rho,t)e^{il\theta}
+ \mathrm{c.c.},
 \label{Fourierser}\end{equation}
inserting the result into the system (\ref{GP2DF})-(\ref{genK}) and solving
for $b_l$ yields the nonlocal Bogoliubov equations for   $(u_l,v_l)$:
\begin{equation}
\fl
i\partial_t\left(\begin{array}{c} u_l\\ v_l
\end{array}\right) = J\left(\begin{array}{cc} L_l + gA\Lambda_l^{-1}A & gA\Lambda_l^{-1}A\\
gA\Lambda_l^{-1}A & L_{-l} + gA\Lambda_l^{-1}A
\end{array}\right)\left(\begin{array}{c} u_l\\ v_l
\end{array}\right)\equiv J H_l\left(\begin{array}{c} u_l\\ v_l
\end{array}\right).
\label{linearsys}\end{equation}
Here $J = \mathrm{diag}(1,-1)$, the operators $\Lambda_l^{-1}$ are understood as
acting on the products of $Au_l$ or $Av_l$, and the operators $L_{\pm l}$ are
defined as
\begin{equation}
L_{\pm l} = -\nabla^2_\rho + V(\rho) + \frac{(n\pm l)^2}{\rho^2} + g F_0 -
(\mu + \Omega (n \pm l)).
 \label{defLl}\end{equation}
The  matrix operator $H_l$ introduced by (\ref{linearsys}) is real and
symmetric and  has also the following property
\begin{equation}
\tau H_l \tau = H_{-l},\quad \tau \equiv \left(\begin{array}{lr} 0 & 1\\ 1 & 0
\end{array}\right).
 \label{propHl}\end{equation}
The operator $H_l$ is related to the orbital projection of the Hessian in the
rotating reference frame. Indeed, the solution $\psi = A(\rho)e^{in\theta -
i\mu t}$ is the stationary point of the Lagrange-modified energy functional
in the rotating frame
\begin{equation}
E \equiv \mathcal{E} - \mu N =  \int\mathrm{d}^2{\brho}\left\{ |\nabla\psi|^2
+ V|\psi|^2 - \Omega\psi^*L_z\psi + \frac{g}{2}F|\psi|^2\right\} - \mu N,
 \label{energy}\end{equation}
(i.e., $\delta E/\delta\psi^*  = 0$ at the solution) where $N =
\int\mathrm{d}^2{\brho}|\psi|^2$ is the number of atoms. Equation
(\ref{linearsys}) can be obtained from the following one
\begin{equation}
\fl
i\partial_t\left(\begin{array}{c} \Phi\;\\ \Phi^*
\end{array}\right) = Je^{-in\theta J}\left(\begin{array}{cc} \frac{\delta^2E}{\delta\psi\delta\psi^*} &
\frac{\delta^2E}{\delta\psi^*\delta\psi^*}\\
\frac{\delta^2E}{\delta\psi\delta\psi} & \frac{\delta^2E}{\delta\psi^*\delta\psi}
\end{array}\right)e^{in\theta J}\left(\begin{array}{c} \Phi\;\\ \Phi^*
\end{array}\right)\equiv e^{-in\theta J} H_{ess} e^{in\theta J}\left(\begin{array}{c} \Phi\;\\ \Phi^*
\end{array}\right)
 \label{sysforPhi}\end{equation}
by expansion  into the Fourier series with respect to $\theta$. Using this
fact and the properties of the operator $H_l$, one can write the expansion of
the Lagrange-modified energy function (\ref{energy}) in powers of $\epsilon$
as follows
\[
E = E_0 + \frac{\epsilon^2}{2}\int\mathrm{d}^2{\brho}(\Phi^*,\Phi)e^{-in\theta J}
H_{ess} e^{in\theta J}\left(\begin{array}{c} \Phi\;\\ \Phi^*\end{array}\right) +
{\cal O}(\epsilon^3) \]
\begin{equation}
\quad = E_0 + 2\pi\epsilon^2\sum_{l\ge0}\int\limits_0^\infty\rho\mathrm{d}\rho
(u^*_l,v^*_l)H_l\left(\begin{array}{c} u_l\\ v_l
\end{array}\right)+ {\cal O}(\epsilon^3).
 \label{Eexp}\end{equation}

For stability according to Lyapunov all the terms in the series in equation
(\ref{Eexp}) must be positive. The $n$-fold axial vortex solution is
spectrally unstable if the radial linear problems
\begin{equation}
JH_l\left(\begin{array}{c} X_1\\ X_2
\end{array}\right) = i\sigma\left(\begin{array}{c} X_1\\ X_2
\end{array}\right), \quad  l = 0,1,2,\ldots,
 \label{eigvalpr}\end{equation}
allow eigenvalues $\sigma$ with non-zero real part (the associated solution to the
linear system (\ref{linearsys}) involves the exponent multiplier $e^{\sigma t}$).

It is helpful to remind the general result on stability of the stationary
points of Hamiltonian systems \cite{MacKay}. It is known that ($i$) the
eigenvalues  appear in quartets $\Xi \equiv\{ \sigma, -\sigma^*, -\sigma,
\sigma^*\}$; ($ii$) the second-order correction to  energy taken over the
subspace corresponding to a quartet of eigenvalues with a non-zero real
part has indefinite signature; ($iii$) the stationary point may loose the
spectral stability only by collision of the (imaginary) eigenvalues bearing
the signatures of opposite sign or by collision at zero $\sigma=0$.

To a quartet of  eigenvalues $\{ \sigma, -\sigma^*, -\sigma, \sigma^*\}$, with
$\sigma$ being the eigenvalue from (\ref{eigvalpr}), the following eigenfunctions,
additional to that in (\ref{eigvalpr}), can be associated:
\begin{eqnarray}
\fl JH_l\left(\begin{array}{c} X_1^*\\ X_2^*
\end{array}\right) = -i\sigma^*\left(\begin{array}{c} X_1^*\\ X_2^*
\end{array}\right), \quad JH_{-l}\left(\begin{array}{c} X_2\\ X_1
\end{array}\right) = -i\sigma\left(\begin{array}{c} X_2\\ X_1
\end{array}\right), \nonumber\\
 JH_{-l}\left(\begin{array}{c} X_2^*\\ X_1^*
\end{array}\right) = i\sigma^*\left(\begin{array}{c} X_2^*\\ X_1^*
\end{array}\right).
 \label{eigenfunct}\end{eqnarray}
This is due to the fact that $H_l$ is symmetric real operator possessing
property (\ref{propHl}). It is easy to establish that all the eigenvalues
from a given quartet bear the same signature. The energy corresponding to
the eigenfrequency $\omega$ ($\sigma = -i\omega$) equals to $\omega\langle
X|J|X\rangle$ where $\langle X|J|X\rangle = 2\pi\int_0^\infty
\rho\mathrm{d}\rho (|X_1|^2 - |X_2|^2)$.

For the following it is convenient to get rid off the diagonal terms $-\Omega l$ in
the operator $JH_l$ by expanding it as follows (here $I$ is the unit matrix)
\begin{equation}
JH_l = J\widetilde{H}_l - \Omega l I.
\label{shiftH}\end{equation}
The new  operator $J\widetilde{H}_l$ has the same eigenfunctions, which now
correspond to the shifted eigenvalues: $\tilde{\sigma} \equiv \sigma - i\Omega l$.
The stable orbital numbers $l$ correspond to non-negative  eigenvalues of
$J\widetilde{H}_lJ\widetilde{H}_l$ (i.e., $-\tilde{\sigma}^2\ge0$). [Here we note
that the instability with a polynomial growth in time $t$ is not possible in the
system (\ref{linearsys}), since the two eigenvalues $\sigma$ and $-\sigma$
coinciding at zero belong to different operators, namely $JH_l$ and $JH_{-l}$, see
equation (\ref{eigenfunct}).] We will prove the inequality in the theorem by
showing that the operator $J\widetilde{H}_lJ$ is positive for $|l|\ge 2|n|$,
$l\ne0$ and non-negative for $l=0$ with one zero mode.

Indeed, consider the inner product of $J\widetilde{H}_lJ$ with $X =(X_1,X_2)^T$:
\begin{eqnarray}
\fl \langle X|J\widetilde{H}_lJ|X\rangle =
2\pi\int\limits_0^\infty\rho\mathrm{d}\rho \Bigl\{X_1^*(\tilde{L}_l +
gA\Lambda^{-1}_lA)X_1 + X_2^*(\tilde{L}_{-l}
+gA\Lambda^{-1}_lA)X_2 \nonumber\\
\lo - g(X_1^*gA\Lambda^{-1}_lAX_2 + X_2^*gA\Lambda^{-1}_lAX_1)\Bigr\},
\label{innprod}\end{eqnarray} where $\tilde{L}_{\pm l} \equiv L_{\pm l} \pm
\Omega l = L_0 + {l(l\pm 2n)}\rho^{-2}$. The operator $L_0$ is
non-negative.  The quickest way to see this is to rewrite $L_0$ as follows
(due to equation (\ref{eqAn}))
\begin{equation}
L_0 = - \frac{1}{\rho A}\frac{\mathrm{d}}{\mathrm{d} \rho}\rho
A^2\frac{\mathrm{d}}{\mathrm{d} \rho} \frac{1}{A}
\label{factL0}\end{equation}
and take into account that, in the inner product  defined by the integral
$\int\rho\mathrm{d}\rho(\cdot)$, the r.h.s. in (\ref{factL0}) can be factorized via
the integration by parts. Using non-negativity of the operator $L_0$ and dropping
the following non-negative term  on the r.h.s. of (\ref{innprod})
\begin{eqnarray}
\fl \int\limits_0^\infty\rho\mathrm{d}\rho\left\{X_1^*gA\Lambda^{-1}_lAX_1 +
X_2^*gA\Lambda^{-1}_lAX_2- g(X_1^*gA\Lambda^{-1}_lAX_2 +
X_2^*gA\Lambda^{-1}_lAX_1)\right\} \nonumber\\
\lo = \int\limits_0^\infty\rho\mathrm{d}\rho\left\{(X_1^* -
X_2^*)gA\Lambda_l^{-1}A(X_1 - X_2)\right\} \ge 0,
\label{dropped}\end{eqnarray}
(since $\Lambda^{-1}_l$ is assumed positive in the formulation of the theorem)
 we arrive at the following
inequalities
\begin{eqnarray}
\fl \langle X|J\widetilde{H}_lJ|X\rangle \ge
2\pi\int\limits_0^\infty\rho\mathrm{d}\rho \left\{ \frac{l(l + 2n)}{\rho^2}|X_1|^2
+  \frac{l(l - 2n)}{\rho^2}|X_2|^2\right\}
\nonumber\\
\ge (l^2 - 2|n
l|)2\pi\int\limits_0^\infty\frac{\mathrm{d}\rho}{\rho}(|X_1|^2 + |X_2|^2).
 \label{condit}\end{eqnarray}
[The linear space in which the inner product in inequality (\ref{condit})
exists contains the eigenfunctions. Indeed, $(X_1,X_2)$ represents the
radial part of the eigenfunction
$(X_1(\rho)e^{i(l+n)\theta},X_2(\rho)e^{i(l-n)\theta})$ (see equations
(\ref{sysforPhi}) and (\ref{Fourierser})). Since the latter is regular in
the Euclidean variables $(x,y)$, we conclude that for non-zero $n$ and/or
$l$, $X_1 = {\cal O}(\rho^{|n+l|})$ and $X_2 = {\cal O}(\rho^{|n-l|})$ as
$\rho\to 0$, while for the ground state ($n=0$)  $l=0$, hence the r.h.s. of
(\ref{condit}) is zero.]

Let us establish when the inner product $\langle
X|J\widetilde{H}_lJ|X\rangle$ is zero. Obviously,  $\langle
X|J\widetilde{H}_0J|X\rangle=0$ for  $X_1 = X_2 = A$. For general $l$, the
condition  $X_1=X_2=\kappa A$ ($\kappa$ is a constant multiplier) is
necessary for the dropped terms $X_1^*L_0X_1 +X_2^*L_0X_2$ and the r.h.s.
of equation (\ref{dropped}) to be zero, i.e.,  for the first row in formula
(\ref{condit}) to be related by the equality sign. But then from the r.h.s.
of (\ref{condit}) (the first row) we conclude that $l=0$, i.e., no
additional zero modes are possible. Thus, for $|l|\ge 2|n|$ and $l\ne0$ the
operator $J\widetilde{H}_lJ$ is strictly positive and for $l=0$ it is
non-negative with one zero mode $X = (A,A)^T$. Such is also  the operator
$\widetilde{H}_l$, in which case the zero mode is $(A,-A)^T$. Now the
eigenvalue problem (\ref{eigvalpr}) can be rewritten as follows
\begin{equation}
\widetilde{H}_l\left(\begin{array}{c}X_1\\X_2\end{array}\right) =
-\tilde{\sigma}^2(J\widetilde{H}_lJ)^{-1}\left(\begin{array}{c}X_1\\X_2\end{array}\right)
+ \delta_{l,0}\beta\left(\begin{array}{c} A\\A\end{array}\right),
\end{equation}
where the application of the inverse of $J\widetilde{H}_lJ$ to the vector
$(X_1,X_2)^T$ on the r.h.s. is justified since for $\tilde{\sigma}\ne0$
this eigenvector is orthogonal to $(A,A)^T$ by the Fredholm alternative
theorem. The minimal eigenvalue of $J\widetilde{H}_lJ\widetilde{H}_l$  is
therefore defined as
\begin{equation}
\mathrm{min}(-\tilde{\sigma}^2) = \mathrm{min}\frac{\langle
X|\widetilde{H}_l|X\rangle}{\langle X|(J\widetilde{H}_lJ)^{-1}|X\rangle},
\label{minimiz}\end{equation}
where the quotient is minimized in the orthogonal complement to the direction given
by $(A,A)^T$. Here we remind that for the spectral stability   the quotient on the
r.h.s. of (\ref{minimiz}) must take only non-negative values. But the latter
quotient is positive for $|l|\ge2|n|$ and $l\ne0$. For $l=0$ we have one zero mode
$(A,-A)^T$. Q.E.D.
\medskip

\noindent \textbf{Remark} In the course of the proof we have established
positivity of the operators $\widetilde{H}_l$ (the orbital projections of
the Hessian for $\Omega=0$) with $|l|\ge 2|n|$ and $l=0$ for the $n$-fold
axial vortex solution (the zero mode of the operator $\widetilde{H}_0$
corresponds to the invariance of the system (\ref{GP2DF})-(\ref{genK}) with
respect to a phase shift).

The established positivity property, however, does not imply the thermal or
Lyapunov stability  of the ground state for non-zero $\Omega$, since the
operators $H_l = \widetilde{H}_l -\Omega l J$ are not, in general,
positive.

It is known that there is a threshold value of the rotation frequency above
which the ground state loses its Lyapunov stability against the surface mode
excitations by a perturbation breaking the rotational symmetry  (see
Refs.~\cite{surfmode1,surfatfint}). Indeed, this is a consequence of the
formula (see equation (\ref{shiftH}))
\begin{equation}
\langle X| H_l |X\rangle = (\omega - \Omega l)\langle X|J|X\rangle,
\label{destabiliz}\end{equation} where $\omega \equiv i\sigma$ is the
eigenfrequency of the oscillations with orbital number $l$ in the
non-rotating system, and $X$ is the corresponding eigenfunction. For fixed
$l$ the above instability  threshold  is given by the quotient $\Omega_{l}
= \omega/l$. For a general perturbation, the critical rotation frequency at
which the nonvortex state loses its local stability is thus given by the
formula $\Omega_c = \mathrm{min}(\omega_l/l)$, where the minimum is
determined over the all resonances with the surface modes \cite{surfmode1}.

The Hessian corresponding to the nonvortex state possesses an additional
property. Setting $n=0$ in equation (\ref{defLl}) we get $L_{\pm l} = L_0
+l^2/\rho^2 \mp\Omega l$ giving
\begin{equation}
H_{-l} = H_l + 2\Omega lJ.
 \label{symmHl}\end{equation}
Therefore, for the nonvortex state at zero rotation ($\Omega=0$) the partial
operators  $JH_l$ and $JH_{-l}$ both have complex conjugate pairs of the
eigenvalues: $\sigma = \pm i\omega$.

%%%%%%%%%%%%%%%%%%%%%%%%%%%%%%%%%%%%%%%%%%%%%%%%%%%%%%%%%%%%%%%%%%%%%%%%%%%%%%%%%%%%%%

\section{Numerical results}
\label{Numerics}

We consider the two-dimensional nonlocal GP model
(\ref{Fdef})-(\ref{GP2DF}) with the interaction potential corresponding to
the Lorentzian. In other words, we restrict ourselves to the straight
vortex lines, which corresponds to the oblate geometry - the simplest
possible case. Thus, investigation of the vortex bending and formation of
vortex rings is beyond of the scope of the present paper. Though the
results below are given only for the parabolic external potential, we have
checked that addition of a perturbation to the potential does not affect
the conclusions on the nonlocality-induced effects (see fig 2 below, for
instance). [Similar results on the nonlocality-induced instabilities hold
also in a completely different external potential -- the tangent-shaped
trap, which models the weak confinement. For instance, figures 1, 3, 4, and
5 below have their counterparts for the tangent trap. ]

To study $n$-vortex solutions and their spectra within the one-dimensional
systems (\ref{eqAn})-(\ref{eqF0Lorentz}) and (\ref{linearsys}) in the
radial variable  we have used two pseudospectral methods \cite{PSM} based
on the Fourier and Chebyshev expansions and used the Fourier-based method
in search for the vortex configurations in 2D. For the radial variable we
have used of up to 256 grid points (which is the number of Fourier modes
and the degree of the Chebyshev polynomial), while the 2D grid was
$128\times128$. The eigenvalues and eigenfunctions of the nonlocal linear
problem (\ref{linearsys}) were found by using the LAPACK routines.

To find axial  $n$-vortex solutions and  two-dimensional vortex
configurations we have minimized the generalized Rayleigh functional (the
Rayleigh functional is also known as the Rayleigh quotient in applications
to linear systems). For the GP equation we define the generalized Rayleigh
functional as the energy functional $\mathcal{E}(\psi)$  evaluated at the
normalized function:
\begin{equation}
\fl R(\phi) \equiv \mathcal{E}\left(\frac{\phi}{||\phi||}\right) =
\left.\int\mathrm{d}^2{\brho}\left\{ |\nabla f|^2 + V|f|^2 - \Omega f^*L_z
f + \frac{g}{2}F(f)|f|^2\right\}\right|_{f=\frac{\phi}{||\phi||}}
 \label{RayleighQ}\end{equation}
(recall that our interaction coefficient $g$ is in fact  ``$gN$'' with $N$
being a fixed number of atoms, see equation (\ref{rescaledvar})). Here the
functional $F(\phi)$ is given by equation (\ref{genK}) and $||\phi||^2 =
\int\mathrm{d}^2{\brho}|\phi|^2$.

The most important feature of the minimization method for linear systems
based on the Rayleigh quotient is transferred by our definition of the
Rayleigh functional to the nonlinear models: the numerical iterations
converge \textit{only} to an actual solution of the stationary GP equation.
To show this consider the variation of the functional $R(\phi)$:
\begin{equation}
\frac{\delta R}{\delta\phi^*} = \left.\frac{1}{||\phi||}\left[ \frac{\delta
\mathcal{E}(f)}{\delta f^*} - \mathrm{Re}\left\{\int\ \mathrm{d}^2\brho
f^*\frac{\delta \mathcal{E}(f)}{\delta
f^*}\right\}f\right]\right|_{f=\frac{\phi}{||\phi||}}.
 \label{nablaR}\end{equation}
Now, convergence of the numerical iteration $\phi_n$ to a minimizer
$\phi_\infty$ of the Rayleigh functional means, evidently, that the
variation on the l.h.s in equation (\ref{nablaR}) tends to zero. To see the
connection to the stationary GP equation, note that as $n\to \infty$ we
have
\begin{equation}
 C_n \equiv\mathrm{Re}\left.\left\{\int\ \mathrm{d}^2\brho
f_n^*\frac{\delta \mathcal{E}(f_n)}{\delta
f_n^*}\right\}\right|_{f_n=\frac{\phi_n}{||\phi_n||}}\to C_\infty
 \label{explRF}\end{equation}
(existence of this limit was confirmed numerically). Setting $\mu \equiv
C_\infty$ and $\psi \equiv f_\infty$ (while the actual solution $\psi$ is
normalized to 1, the variable $\phi$ in (\ref{nablaR}) is not) we obtain
the solution to the stationary GP equation, which can be given  also as
$\mu\psi = {\delta \mathcal{E}(\psi)}/{\delta \psi^*}$, since in the limit
$n\to\infty$ equation (\ref{nablaR})  dictates
\begin{equation}
\frac{\delta \mathcal{E}(f_\infty)}{\delta f_\infty^*} - C_\infty f_\infty
= 0.
\end{equation}
We emphasize that in our numerical scheme the chemical potential $\mu$ is
computed  {\it alongside} with computation of the stationary solution
$\psi$, namely by the iteration: $C_n\to \mu$ as $n\to \infty$. That is why
only actual solutions of the stationary GP equation can be  minimizers of
the Rayleigh functional.

Moreover, it is easy to see that the Rayleigh quotient is also the Lyapunov
functional for the local minima (for fixed number of atoms), since the
second variation of $R(\phi)$ is given by the second variation of
$\mathcal{E}(f)$ with the substitution $\delta f = \delta(\phi/||\phi||)$.
[It is seen that the method based on the generalized Rayleigh quotient is
ideally suited  for application to computation of the stationary solutions
of nonlinear systems, it will be given in more detail in a separate
publication \cite{MTHR}.]

We have searched for the energy minimizers by starting from the
topologically different initial functions $\phi_0$ in the minimization
procedure and comparing the energy of the resulting stationary solutions.
The numerical step $\phi_n\to\phi_{n+1}$ was performed in the direction
opposite to the current gradient $\delta R/\delta\phi_n$, thus ensuring
minimization of the energy at each step. We have observed the fast
convergence of the numerical scheme to the stationary solutions of the GP
equation (we stopped the numerical iterations when the norm of variation
$\delta R/\delta\phi$ and the norms of the  differences $\phi_n-\phi_{n-1}$
and $C_n - C_{n-1}$ were on the order of $10^{-12}$). Finally, we have
checked the accuracy of the grid by varying the grid size and observing the
changes in the solution, in all cases the error of the numerical solution
had negligible effect on the results.

Let us start with the  nonvortex state. One of the characteristic
parameters of the nonvortex state is the rotation frequency $\Omega_v$
above which it ceases to be the global minimum of the energy functional (in
the rotating reference frame).  The threshold rotation frequency was
numerically computed by comparing the energies of the nonvortex state and
the  1-vortex in the rotating frame. The result is presented in figure~1(a)
(in our dimensionless system the rotation frequency is measured in halves
of the trap frequency, see equation (\ref{rescaledvar})). It is seen that
even an unjustifiably large range of interactions only slightly affects the
threshold rotation frequency $\Omega_v$.

We found that the 1-vortex solution has only imaginary eigenvalues $\sigma
= -i\omega$ for all $a\le1$, thus it is dynamically stable in the nonlocal
GP model (\ref{Fdef})-(\ref{GP2DF}). It turns out that, similar as in the
local model, it has one anomalous mode (for $\Omega =0$), i.e., the
eigenvalue corresponding to the orbital operator $H_l$, with $l=1$, in
equation (\ref{linearsys})  and bearing a negative energy. The frequency of
the anomalous mode decreases with increase of the interaction range $a$,
see figure~1(b) (here and below, we chose to show the modes of the
operators $H_l$ with $l>0$, thus our ``anomalous mode'' has negative norm
and positive frequency and is related by equation (\ref{eigenfunct}) to the
true anomalous mode of the operator $H_{-l}$).  It is known that the
frequency of the anomalous mode gives the frequency of vortex precession
around the axis of the trap \cite{anomalmode}. As it was mentioned above,
the results on the nonlocality-induced effects do not depend on the
external trap. To illustrate this we give the analog of figure~1(b) for the
perturbed parabolic trap in figure~2. The 1-vortex solution suffers from a
slight deformation  with respect to variation of the interaction range as
is seen from figure~3.

\begin{figure}
\begin{center}
\includegraphics{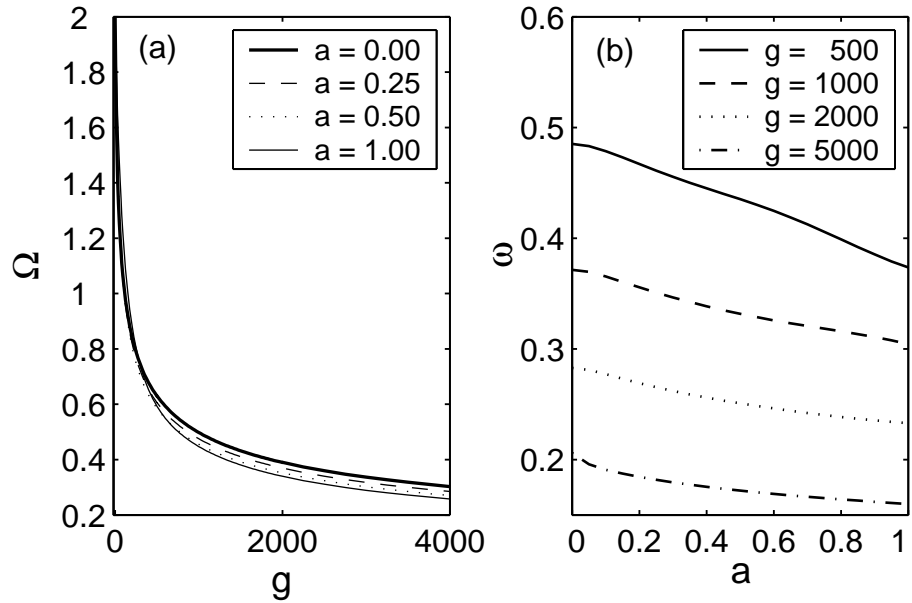}
\caption{The threshold rotation frequency above which the 1-vortex has
lower energy (in the rotating frame) than the nonvortex state vs. the
strength of interaction for various values of the interaction range, panel
(a), and the frequency of the anomalous mode (for $\Omega = 0$) of the
1-vortex vs. the interaction range for various values of the interaction
strength, panel (b).}
\end{center}
\end{figure}

\begin{figure}
\begin{center}
\includegraphics{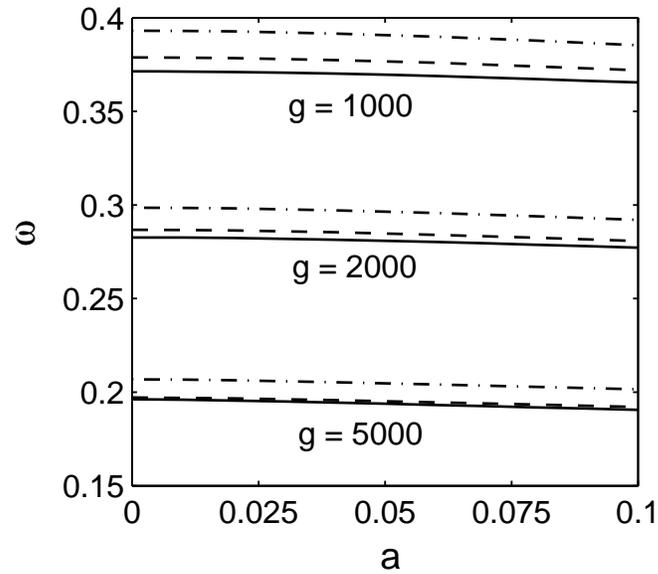}
\caption{The frequency of the anomalous mode (for $\Omega = 0$) of the
1-vortex vs. the interaction range for various values of the interaction
strength for the perturbed harmonic trap (the parabolic potential perturbed
by the quartic term), the dash and dash-dot lines. The solid line
corresponds to the harmonic trap. }
\end{center}
\end{figure}

\begin{figure}
\begin{center}
\includegraphics{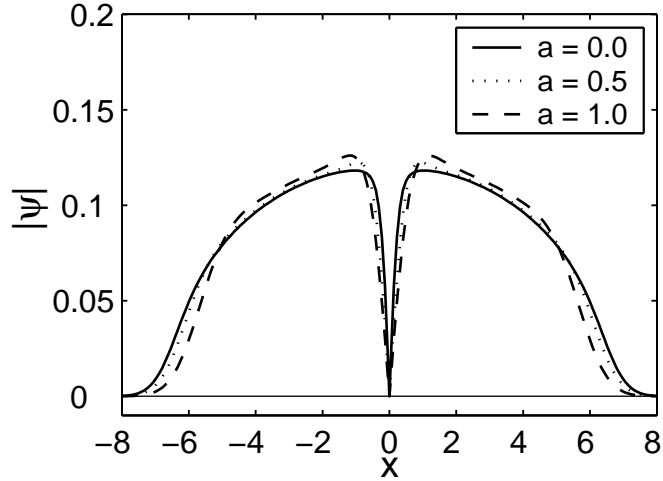}
\caption{The shape of the 1-vortex solution for local, $a=0$, and nonlocal,
$a=0.5; 1$, interactions. The interaction strength is $g = 3000$.}
\end{center}
\end{figure}

We  now turn to the higher axial $n$-fold vortices.  For the local GP
equation it is known that the intervals of the interaction strength $g$
where the axial $n$-vortex is dynamically unstable are intertwined with the
intervals of dynamical stability \cite{AxialStab}. According to the main
result of section~\ref{Stability}, the axial $n$-vortex solution may be
unstable only with respect to the orbital perturbations $\sim e^{il\theta}$
satisfying $0< l< 2n$ (without loss of generality we can consider only
positive $n$ and $l$). The instabilities of the $n$-vortices result in
vortex splitting. This fact can be seen from the Taylor expansion with
respect to $z = x+iy$ and $z^*$ of the perturbed vortex solution with a
perturbation having nonzero projection on the orbital number $l$ (see
formulae (\ref{perturba})-(\ref{Fourierser})) --  for $0<l<2n$ the lowest
power in $z$ and/or $z^*$, which defines the vorticity at the origin, will
be $|n-l|$, i.e., will come from the perturbation.

Variation of the interaction range  $a$  for any fixed interaction strength
$g$ (in physical terms, the interaction strength multiplied by the number
of atoms)  results in appearance of a finite number of instability
intervals, now with respect to the interaction range, due to resonances of
the anomalous and normal modes. We have not found resonances with the
orbital number $l=1$ (which are not forbidden). The anomalous mode with
orbital number $l=1$, which is present for all axial $n$-vortices, has the
closest to zero frequency for any range $a$ ($\Omega = 0$). For the
2-vortex solution the instability intervals vs. the interaction range are
illustrated in figure~4. We found that with increase of the interaction
strength $g$ the instability intervals move towards smaller values of the
interaction range.

\begin{figure}
\begin{center}
\includegraphics{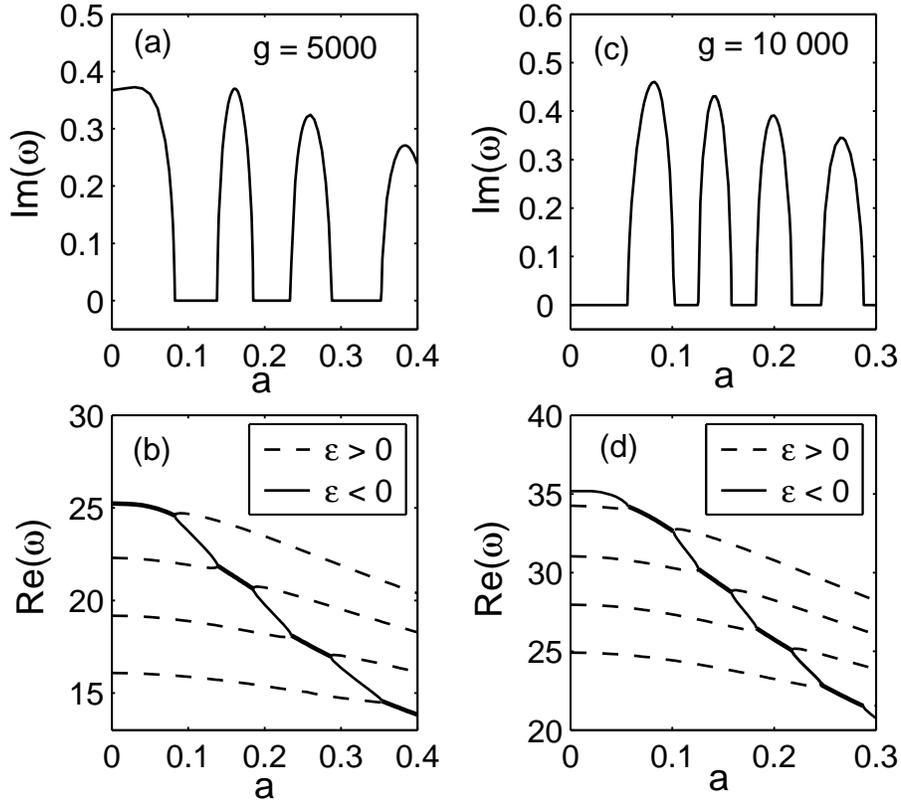}
\caption{The real, (a) and (c), and imaginary, (b) and (d), parts of the
eigenfrequencies corresponding to the orbital linear modes with $l=2$ of
the axial 2-vortex (here $\Omega = 0$). Only the anomalous mode and the
resonant normal modes are shown (the inserts give the signature). The
2-vortex solution loses its stability due to collision resonance of the
anomalous mode (solid line) and the normal modes (dashed lines). }
\end{center}
\end{figure}

Nonlocality is also responsible for appearance of additional anomalous
modes with the orbital numbers higher than the vorticity ($n<l<2n$). For
$\Omega=0$  this happens for the axial $n$-vortices with $n\ge2$, which
have more than one orbital number $l$ in the interval $0< l< 2n$, where the
operators $H_l$ introduced in (\ref{linearsys}) may have negative
eigenvalues. For the 2-vortex solution the additional anomalous mode has
orbital number $l=3$, see figure~5. For the 3-vortex solution, for
instance, we have observed creation of such anomalous mode with  $l=5$. It
should be noted that for some values of the interaction strength there are
anomalous modes with higher orbital numbers in the local model as well. For
instance, the axial 2-vortex solution has anomalous mode with $l=3$ for
$g=2000$. For the 3-vortex solution and $g = 10^4$ there is anomalous mode
with $l=4$ (it resonantly collides with  normal modes  giving rise to
instabilities, see below).

The frequency of the created anomalous mode remains close to zero,
therefore no resonances are possible with further increase of $a$. The
threshold value of $a$ at which the zero  crossing occurs decreases with
increase of the interaction strength, for instance, for $g=10^4$ the
threshold is $a \approx 0.3$ as compared to $a \approx 0.35$ for $g=5\times
10^3$ in figure~5.

\begin{figure}
\begin{center}
\includegraphics{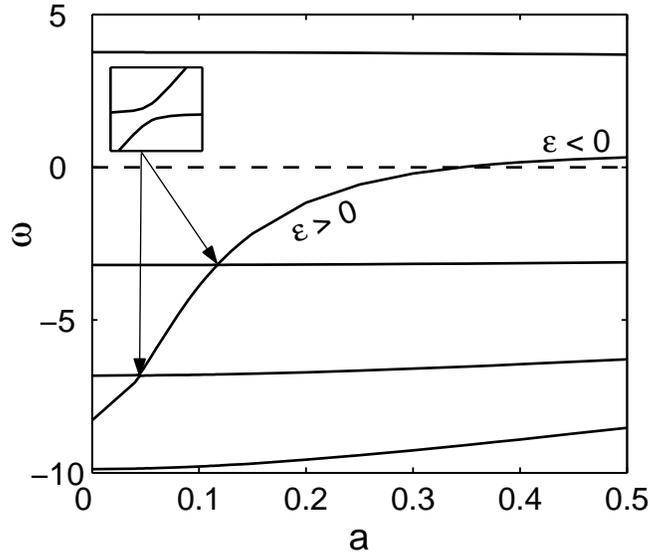}
\caption{Zero crossing with creation of an anomalous mode with the orbital
number $l=3$  of the axial 2-vortex. In the picture $g=5000$ and $\Omega =
0$. The insert indicates the avoided crossing collisions of the normal modes.
}
\end{center}
\end{figure}

Finally, we have found that nonlocality opens new splitting channels for
the axial $n$-vortex in the interval $n<l<2n$. In contrast, as noted in
Ref.~\cite{AxialStab}, the instability channels of the axial $n$-vortices
in the local GP equation satisfy $l\le n$ with the symmetric splitting
instability, $l=n$, being  the strongest. The situation is different in the
nonlocal model since there is an additional parameter -- the range $a$. For
instance, it is possible to find such values of the interaction strength
and the range that the only dynamical instability of the axial $n$-vortex
has the orbital number higher than $n$. This happens already in the case of
the 3-vortex, as seen from figure~6. However, the instabilities with the
orbital numbers $l>n$ turn out to be weak. Similar, for the 4-vortex
solution,  we have found that for $a$ close to 0.1 the only unstable
orbital mode has the orbital number $l=6$ with Re$(\omega)$ of order
$10^{-3}$.

\begin{figure}
\begin{center}
\includegraphics{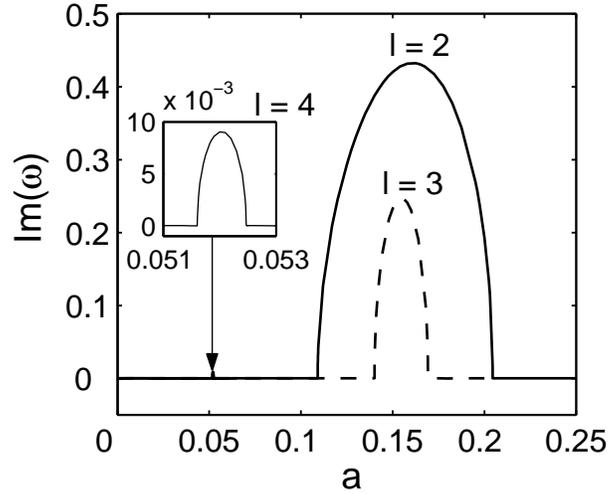}
\caption{The  imaginary part of the eigenfrequencies corresponding to the
unstable orbital  modes with the orbital numbers $l=2,3$ and $l=4$ (in the
insert) of the axial 3-vortex. Here the interaction strength is $g=10^4$.}
\end{center}
\end{figure}

The new channels of instability with $n<l<2n$  seem to suggest existence of
new non-axial vortex solutions, involving combinations of vortices and
antivortices, which could minimize the energy for some  rotation frequency.
However, it turns out that in the nonlocal GP equation, similar as in the
local one, the stationary vortex solutions involving combinations of
vortices and antivortices are never the energy minimizers -- there always
exist another stationary solution with the same total vorticity, but
comprised of vortices only, which minimizes the energy. For instance, the
instability with $l=4$ of the axial 3-vortex solution, shown in figure~6,
could result in the formation of a non-axial 5-vortex solution with 4
vortices and 1 antivortex. Such solution was indeed found by the
minimization starting from the seed resembling the axial  $3$-vortex, see
the right panel of figure~7. For rotation frequencies $\Omega> 0.38$ this
solution  has lower energy than the ground state, however, there is at
least one another solution, the combination of three vortices, shown in the
left panel of figure~7, which has lower energy then the vortex-antivortex
solution for all rotation frequencies. It is interesting to note that both
these solutions do exist in the local model ($a=0$).

\begin{figure}
\begin{center}
\includegraphics{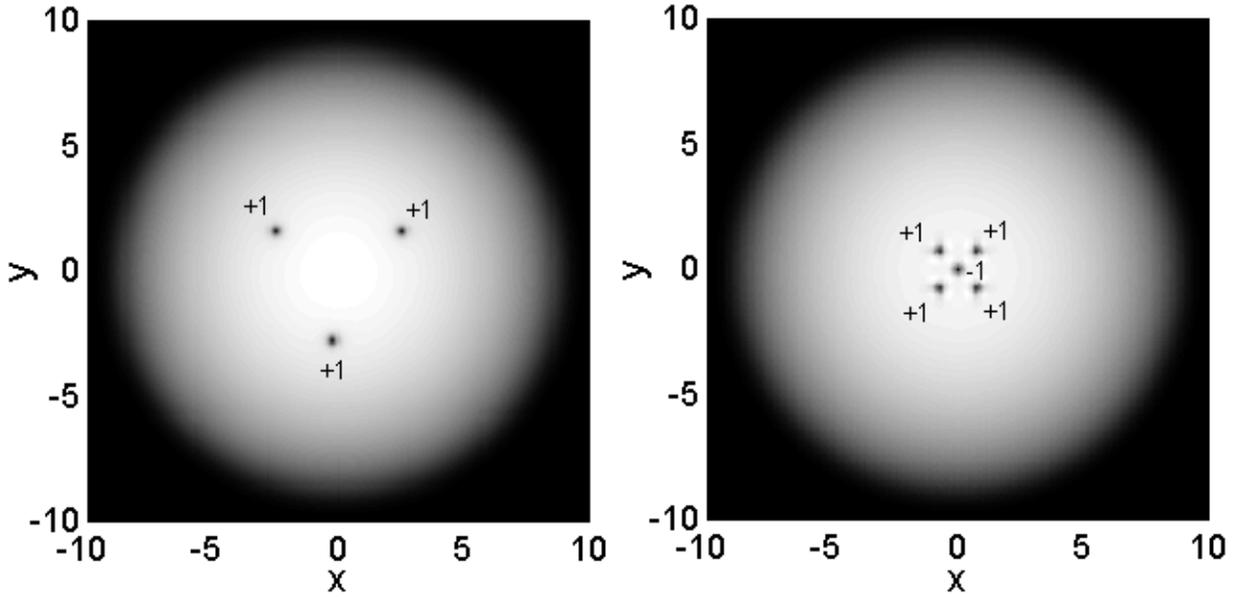}
\caption{The stationary vortex solutions for $a=0.052$, $g=10^4$ and
$\Omega = 0.38$. The left panel shows the 3-vortex solution  and  the right
one -- the 5-vortex solution comprised of 4 vortices and 1 antivortex (in
the center). }
\end{center}
\end{figure}

More complicated combinations of vortices and antivortices were found by
the energy minimization. However, our numerical results indicate  that for
a stationary solution involving vortices and antivortices there always
exist another stationary solution with the same total vorticity, but
comprised of vortices only, which has lower energy. This is true for the
local and nonlocal GP equations. For instance, in figure~8 we give such
solutions with the total vorticity 8: the left panel shows the 8-vortex
solution (the energy minimizer)  and the right one -- the 16-vortex
solution with 12 vortices and 4 antivortices.

\begin{figure}
\begin{center}
\includegraphics{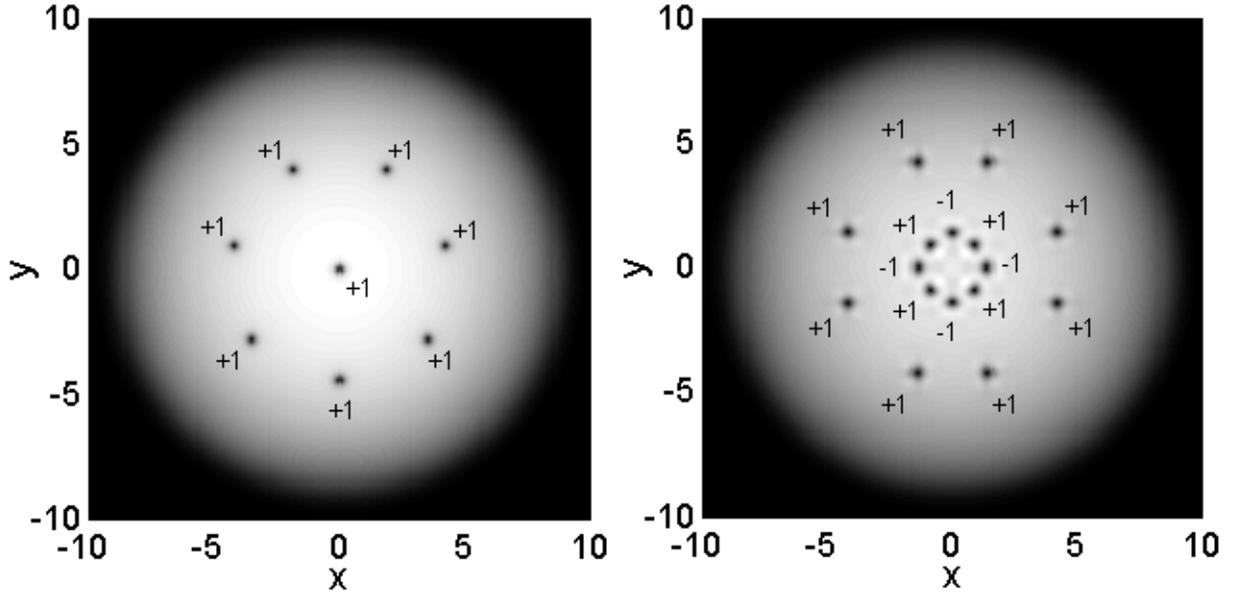}
\caption{The stationary vortex solutions for $a=0.052$, $g=10^4$ and
$\Omega = 0.5$. The left panel shows the 8-vortex solution  and the right
one -- the 16-vortex solution comprised of 12 vortices and 4 antivortices.
}
\end{center}
\end{figure}

\section{Conclusion}
\label{Summary}

In the present paper we have made an attempt to understand what changes in
the properties of quantized vortices can be attributed to the range of
interaction, when the latter is not negligible. The motivation was
similarity of the properties of quantized vortices in the gaseous BECs,
where the range of interactions is much smaller than the size of the vortex
core and, hence, negligible, and in liquid helium, where the  vortex core
size is comparable to the interaction range.

In general, the local GP equation cannot not be justified for description
of the quantized vortices in liquid helium. It is common approach to use a
nonlocal model for the phenomenological description of collective phenomena
in such systems. However, there is only one local model and infinitely many
nonlocal ones, if the derivation from the first principles is difficult or
not possible to carry out. To choose between different nonlocal models one
is left to rely on the analysis of each term in the model to understand the
effect of it. In this paper we have analyzed solely the effect of the
finite range of interaction on the properties of quantized vortices. The
 interaction range affects the stability properties of the $n$-fold
vortices creating new channels for vortex splitting. However it has only
slight effect on the threshold frequency above which the nonvortex state
ceases to be the global energy minimum.

Here we note the difference in the effect of nonlocality in two-dimensional
and one-dimensional GP equations. In 2D  there is always a threshold value
of the interaction range for  appearance of the instabilities caused solely
by nonlocality, while, as it is suggested by numerical simulations in
Ref.~\cite{DK}, in 1D such instabilities may set in beyond all orders of
the range parameter.

Finally, in spite of the fact that the finite interaction range opens new
splitting channels with creation of antivortices, for any value of the
interaction range and any rotation frequency for the vortex-antivortex
solution there is a plain vortex solution of the same vorticity but with a
lower energy. Thus, the annihilation of a vortex-antivortex pair lowers the
energy, which will prevent the vortex-antivortex states from being
observed. However, {\it a priori}, this is not evident, since there is
mutual interaction between the individual vortices which depends on the
model and is not know in detail at a small distance. In the local GP
equation, for instance, the interaction potential is known only at a large
distance, and it has only the logarithmic falloff according to the
Kirchoff-Onsager formula. We can explain the insensitivity of the
topological structure of the minimizers to the interaction range if we
suppose that the leading term in the interaction potential is due to the
phase overlap in the kinetic term (which, in fact, gives nothing but the
Kirchoff-Onsager formula).

In connection to this, it would be interesting to find out what changes to
the vortex interaction will bring the inclusion of a nonlinear and/or
nonlocal correction to the kinetic energy in the density functional. Such
corrections have been suggested in the literature on liquid helium (see,
for instance, \cite{FHe3}). There will be significant changes in the
structure of the energy minimizers as is suggested by a similar but
completely integrable model which also admits  vortex solutions -- the
Euclidean complex sine-Gordon equation. In the latter model the kinetic
energy term is nonlinear, which results in the energy degeneracy with
respect to the number of vortices and antivortices as long as the total
vorticity is fixed \cite{SGeq}. This is a direction for future studies.

\ack

This research was supported by the FAPESP grant. The authors are grateful to
A.~M.~Kamchatnov for helpful  discussions.


\section*{References}
\begin{thebibliography}{99}

\bibitem{gross}  Gross E P 1961 {\it Nuovo Cimento} \textbf{20} 454

\bibitem{pitaevskii} Pitaevskii L P 1961 {\it Zh. Eksp. Teor. Fiz.} \textbf{40} 646  [{\it Sov. Phys.
JETP} \textbf{13} 451]

\bibitem{revBEC1}  Dalfovo F,  Giorgini S, Pitaevskii L P and Stringari S 1999 {\it Rev. Mod. Phys.}
\textbf{71} 463

%%%%%%%%%%%%%%%%%%% VALIDATION OF GROSS-PITAEVSKII THEORY %%%%%%%%%%%%%%%%%%%%%%%%%%%%%%%%%%
\bibitem{BPtheor} Baym G and Pethick C J 1996 {\it Phys. Rev. Lett.} \textbf{76} 6

\bibitem{expannum}  Holland M and  Cooper J 1996 {\it Phys. Rev. A } \textbf{53} R1954

\bibitem{exp1}  Anderson  M H,  Ensher J R,  Matthews M R,  Wieman C E
and Cornell E A 1995 {\it Science} \textbf{269} 198

\bibitem{excitnum} Edwards M, Ruprecht P A, Burnett K, Dodd R J and  Clark C W 1996
{\it Phys. Rev. Lett.} \textbf{77} 1671

\bibitem{excitexp}  Jin D S,  Ensher J R,  Matthews M R,  Wieman C E and  Cornell E A 1996
{\it Phys. Rev. Lett.} \textbf{77} 420

\bibitem{interexp} Andrews M R, Townsend C G, Miesner H-J, Durfee D S, Kurn D M and Ketterle W {\it Science} 1997
\textbf{275} 637

\bibitem{interth} R\"ohrl A, Naraschewski M, Schenzle A and Wallis H 1997 {\it Phys. Rev. Lett.}
\textbf{78} 4143

%%%%%%%%%%% FIRST OBSERVATIONS OF VORTICES %%%%%%%%%%%%%%%%

\bibitem{expvrt1}  Matthews M R,  Anderson B P,  Haljan P C,  Hall D S,  Wieman C E and
 Cornell E A 1999 {\it Phys. Rev Lett. } \textbf{83} 2498

\bibitem{expvrt2}  Madison K W,  Chevy F, Wohlleben W and Dalibard J 2000 {\it Phys. Rev. Lett.} \textbf{84} 806;
 Chevy F,  Madison K W and  Dalibard J 2000 {\it Phys. Rev. Lett.} \textbf{85} 2223

\bibitem{expvrt3}  Abo-Shaeer J R,  Raman C,  Vogels J M and  Ketterle W 2001 {\it Science} \textbf{292} 476

\bibitem{expvrt3a}  Raman C,  Abo-Shaeer J R, Vogels J M, K. Xu and  Ketterle W 2001 {\it Phys. Rev. Lett.}
\textbf{87} 210402

\bibitem{solitvrt1} Anderson B P, Haljan P C, Regal C A, Feder D L,  Collins L A, Clark C W and Cornell E A
2001 {\it Phys. Rev. Lett.} \textbf{86} 2926

\bibitem{solitvrt2} Dutton Z, Buddle M, Slowe C and Hau L V 2001 {\it Science} \textbf{293} 663

\bibitem{expvrt4}  Hodby E,  Hechenblaikner G,  Hopkins S A, Marag\`{o}  O M and  Foot C J 2002
{\it Phys. Rev. Lett.} \textbf{88} 010405

\bibitem{expvrt5} Haljan P C, Coddington I, Engels P and  Cornell E A 2001 {\it Phys. Rev. Lett.} \textbf{87} 210403

\bibitem{coreless1}  Leanhardt A E, Shin Y, Kielpinski D, Pritchard D E, and  Ketterle W 2003 {\it Phys. Rev. Lett.}
\textbf{90} 140403


%%%%%%%%%%%%%% NUCLEATION OF VORTICES IN BECs %%%%%%%%%%%

\bibitem{OmegEner} Lundh E, Pethick C J and Smith H 1997 {\it Phys. Rev. A} \textbf{55} 2126

\bibitem{phasdiagr} Isoshima T and Machida K 1999 {\it  Phys. Rev. A} \textbf{60} 3313

\bibitem{anomalmode} Svidzinsky A A and  Fetter A L 2000 {\it Phys. Rev. Lett.} \textbf{84} 5919;
Feder D L, Svidzinsky A A, Fetter A L and Clark  C W 2001 {\it
Phys. Rev. Lett.} \textbf{86} 564

\bibitem{surfmode1} Dalfovo F and Stringari S 2001 {\it Phys. Rev. A} \textbf{63} 011601

\bibitem{vrtnuclcloud1} Williams J E, Zaremba E, Jackson B, Nikuni T and Griffin A 2002 {\it Phys. Rev. Lett.}
\textbf{88} 070401

\bibitem{thermcloud}  Penckwitt A A, Ballagh R J and Gardiner C W  2002 {\it Phys. Rev. Lett.} \textbf{89} 260402

\bibitem{stabdiagrfint} Mizushima T, Isoshima T and Machida K 2001 {\it Phys. Rev. A} \textbf{64} 043610

\bibitem{surfatfint} Simula T P, Virtanen S M M and Salomaa M M 2002 {\it Phys. Rev. A} \textbf{66} 035601

\bibitem{numer1}  Garc\'ia-Ripoll J J and P\'erez-Garc\'ia  V M 1999 {\it Phys. Rev. A} \textbf{60} 4864

\bibitem{numer2}  Garc\'ia-Ripoll J J and P\'erez-Garc\'ia  V M 2001 {\it Phys. Rev. A} \textbf{63} 041603


\bibitem{localnucl1} Crescimanno M, Koay C G, Peterson R and Walsworth R 2000 {\it Phys. Rev. A}
\textbf{62} 063612

\bibitem{localnucl2} Lundh E, Martikainen J P and Suominen Kalle-Antti 2003 {\it Phys. Rev. A}
\textbf{67} 063604

\bibitem{vortform} Jackson B, McCann J F, and Adams C S 1998 {\it Phys. Rev. Lett.} \textbf{80} 3903

\bibitem{vortshed}  Winiecki T, Jackson B, McCann J F, and Adams C S 2000 {\it J. Phys. B}  \textbf{33} 4069

\bibitem{vortring} Winiecki T and Adams C S 2000 {\it Europhysics Lett.} \textbf{52} 257


%%%%%%%%%%%%%%%%%%%%% DERIVATION OF THE GROSS-PITAEVSKII EQUATION %%%%%%%%%%%%

\bibitem{deriv3D} Lieb  E H, Seiringer R and Yngvason J 2000 {\it Phys. Rev. A} \textbf{61} 043602

\bibitem{deriv2D} Lieb  E H, Seiringer R and Yngvason J 2001 Commun. Math. Phys. \textbf{224} 17

\bibitem{derivnew}  Lieb E H and  Seiringer R 2002 {\it Phys. Rev. Lett.} \textbf{88} 170409

%%%%%%%%%%%%%%%%%%%%%%%%%% HELIUM II %%%%%%%%%%%%%%%%%%%%%%%%%%%%%%%%%%%%%%%%%%%%%%%%%%%%%%

\bibitem{OnsgFeynm} Onsager L 1949 {\it Nuovo Cimento} \textbf{6}, Suppl. 2, 249 and 281;
Feynman R P 1955  In {\it Progress in Low Temperature Physics}
Vol. I p. 17 ed. Gorter C J (Amsterdam: North Holland Publ.)

\bibitem{vrtHe}  Donnelly R J 1991 {\it Quntized Vortices in
Helium II} (Cambridge, UK: Cambridge University Press)

\bibitem{vrtdyn} Saffman  P G 1992 {\it Vortex Dynamics} (Cambridge, UK:  Cambridge University Press)

\bibitem{hist}  Balibar S 2003 {\it Looking back at superfluid helium} e-print cond-mat/0303561

\bibitem{Vinen}  Vinen W F 1961 {\it Proc. Roy. Soc. A} \textbf{240} 114

\bibitem{photvrtHe1}  Williams G A and  Packard R E 1974 {\it Phys. Rev. Lett.} \textbf{33} 280

\bibitem{photvrtHe2} Yarmchuck E J and  Packard R E 1982 {\it J. Low Temp. Phys.} \textbf{46} 479

%%%%%%%%%%%%%%%%%% FUNCTIONAL APPROACH FOR HELIUM II %%%%%%%%%%%%%%%%%%%%%%%%%%%%%%%%%%%%%%%%%

\bibitem{FHerev} Dalfovo F and  Stringari S 2001 {\it J. Chem. Phys.} \textbf{115} 10078

\bibitem{FHe1} Stringari S and Treiner J 1987 {\it Phys. Rev. B} \textbf{36} 8369; 1987
{\it J. Chem. Phys.} \textbf{87} 5021

\bibitem{FHe2}  Dupont-Roc J,  Himbert M, Pavloff N and Treiner J 1990 {\it J. Low Temp. Phys.} \textbf{81} 31

\bibitem{newhe1}  Dalfovo F 1992 {\it Phys. Rev. B} \textbf{46} 5482

\bibitem{FHe3}  Dalfovo F,  Lastri A,  Pricaupenko L,  Stringari S  and Treiner J 1995 {\it Phys. Rev. B}
\textbf{52} 1193

\bibitem{FHe4}  Berloff N G and  Roberts P H 1999 {\it J. Phys. A: Math. Gen.} \textbf{32} 5611

%%%%%%%%%%%%%%%%%%%%%% VORTICES IN HELIUM II AND THE LOCAL THEORY %%%%%%%%%%%%%%%%%%%%%%%%%%%%%%%%%%%%

\bibitem{highorder}  Berloff N G 1999 {\it J. Low Temp. Phys.} \textbf{116} 359

\bibitem{Heloc1}  Jones C A and  Roberts P H 1982 {\it J. Phys. A: Math. Gen.}  \textbf{15} 2599

\bibitem{Heloc2} Frisch T, Pomeau Y and Rica S 1992 {\it Phys. Rev. Lett.} \textbf{69} 1644

\bibitem{Heloc3} Koplik J and  Levine H 1993 {\it Phys. Rev. Lett.} \textbf{71} 1375;
1996 {\it Phys. Rev. Lett.} \textbf{76} 4745

\bibitem{Heloc4} Nore C,  Brachet M E and  Fauve S 1993 {\it Physica D} \textbf{65} 154

\bibitem{FHe5} Berloff N G and Roberts P H 2000 {\it Physics Letters A} \textbf{274} 69
%%%%%%%%%%%%%%%%%%%%%%%%%%% NONLOCALITY IN GROSS-PITAEVSKII THEORY %%%%%%%%%%%%%%%%%%%%%%%%%%%

\bibitem{negnonloc1}  P\'erez-Garc\'ia V M,  Konotop V V and  Garc\'ia-Ripoll J J 2000
{\it Phys. Rev. E} \textbf{62} 4300

\bibitem{negnonloc2} Parola A, Salasnich L and Reatto L 1998 {\it  Phys. Rev. A} \textbf{57} R3180

\bibitem{negnonloc3} Salasnich L 1999 {\it Phys. Rev. A}  \textbf{61} 015601

%%%%%%%%%%%%%%%%%%%%%%%%%%%%%% BESSEL FUNCTIONS %%%%%%%%%%%%%%%%%%%%%%%%%%%%%%%%%%%%%

\bibitem{Bessel}  Lebedev N N 1972 {\it Special functions and their applications}
(New York: Dover Publications)

%%%%%%%%%%%%%%%%%%%%%%%%%%%%%% STABILITY OF AXIAL VORTICES %%%%%%%%%%%%%%%%%%%%%%%%%%%%%%%

\bibitem{stab1}  Svidzinsky A A  and  Fetter A L 2000 {\it Phys. Rev. Lett.} \textbf{84} 5919;
1998 {\it Phys. Rev. A} \textbf{58} 3168

\bibitem{stab2} Rokhsar D 1997 {\it Phys. Rev. Lett.} \textbf{79} 2164

\bibitem{anmmode}  Dodd R J, Burnett K, Edwards M and Clark C W 1997 {\it Phys. Rev. A}  \textbf{56} 587


\bibitem{rolenegenr}  Skryabin D V 2000 {\it Phys. Rev. A} \textbf{63} 013602

\bibitem{AxialStab}  Pu H,  Law C K, Eberly J H and  Bigelow N P 1999 {\it Phys. Rev. A} \textbf{59} 1533

\bibitem{attrinstab} Saito H and Ueda M 2002 {\it Phys. Rev. Lett.} \textbf{89} 190402

%%%%%%%%%%%%
\bibitem{stabGL2D} Aranson I S and Steinberg V 1996 {\it  Phys. Rev. B} \textbf{53} 75

\bibitem{stabGL3D1} Gabbay M, Ott E and  Guzdar P N 1997 {\it  Phys. Rev. Lett.} \textbf{78} 2012

\bibitem{stabGL3D2}  Aranson I S and  Bishop A R 1997 {\it  Phys. Rev. Lett.} \textbf{79} 4174

\bibitem{optvrtstab}  Kivshar Yu S and  Luther-Davis B 1998 {\it Phys. Rep.} \textbf{298} 81

\bibitem{MacKay}  Mackay R S 1987 {\it Hamiltonian Dynamical Systems} eds. Mackay R S and Meiss J D
(Bristol: Hilger) p 137
%%%%%%%%%%%%%%%%%%%%

\bibitem{PSM} For introduction to the pseudospectral methods consult:

Fornberg B 1996 \textit{A Practical Guide to Pseudospectral Methods}
(Cambridge, UK: Cambridge University Press);  Boyd J P 2000 \textit{Chebyshev
and Fourier Spectral Methods} Second Edition (New York: DOVER Publications
Inc.); Trefethen L N 2000 \textit{Spectral Methods in Matlab} (Philadelphia:
SIAM)
%%%%%%%%%%%%%%%%%

\bibitem{MTHR} Shchesnovich V S  {\it The generalized Rayleigh quotient for nonlinear systems} to be published

\bibitem{DK} Deconinck  B and Kutz J N 2002 {\it Phys. Lett. A} \textbf{319}, 97 (2003)

\bibitem{SGeq} I. V. Barashenkov, V. S. Shchesnovich, and R. Adams, {\it Nonlinearity} \textbf{15}, 2121 (2002)



\end{thebibliography}
\end{document}